# Polyacrylonitrile/Graphene Nanocomposite: Towards the Next Generation of Carbon Fibers

Siavash Rajabpour[1], Qian Mao[2], Zan Gao[3], Mahdi Khajeh Talkhoncheh[1], Jiadeng Zhu[3], Yosyp Schwab[3], Malgorzata Kowalik[2], Xiaodong Li[3], Adri C.T. van Duin[1,2]

[1]Department of Chemical Engineering, The Pennsylvania State University, University Park, Pennsylvania 16802, United States

[2]Department of Mechanical Engineering, The Pennsylvania State University, University Park, Pennsylvania 16802, United States

[3]Department of Mechanical and Aerospace Engineering, University of Virginia, 122 Engineer's Way, Charlottesville, Virginia 22904, United States

*Corresponding authors. Emails: acv13@psu.edu (Adri C.T. van Duin); xl3p@virginia.edu (Xiaodong Li)

**Abstract**

Carbon Fibers (CFs) are the key solution for the future lightweight vehicle with enhanced fuel efficiency and reduced emissions owing to their ultrahigh strength to weight ratio. However, the high cost of the current dominant polyacrylonitrile (PAN)-based CFs hinders their application. The use of low-cost alternative precursors may overcome this issue. Unfortunately, low-cost CFs derived from cheaper single component precursors suffer from poor mechanical properties. Developing composite CFs by adding nanoadditives is very promising for low-cost CFs. Therefore, a fundamental understanding of carbonization condition impacts and polymer/additives conversion mechanisms during whole CF production are essential to develop low-cost CFs. In this work, we have demonstrated how the carbonization temperature affects the PAN/graphene CFs properties by performing a series of ReaxFF based molecular dynamics simulations. We found that graphene edges along with the nitrogen and oxygen functional groups have a catalytic role and act as seeds for the graphitic structure growth. Our MD simulations unveil that the addition of the graphene to PAN precursor modifies all-carbon membered rings in the CFs and enhances the alignments of 6-member carbon rings in carbonization which leads to superior mechanical properties compare to PAN-based CFs. These ReaxFF simulation results are validated by experimental structural and mechanical characterizations. Interestingly, mechanical characterizations indicate that PAN/graphene CFs carbonized at 1250 ºC demonstrate 90.9 % increase in strength and 101.9% enhancement in Young's modulus compare to the PAN-based CFs carbonized at 1500 ºC. The superior mechanical properties of PAN/graphene CFs at lower carbonization temperatures offers a path to both energy savings and cost reduction by decreasing the carbonization temperature and could provide key insights for the development of low-cost CFs.

## 1. Introduction

Since the commercialization of carbon fibers (CFs) in the 1960s, the remarkable properties of CFs such as high modulus and tensile strength, low density and notable chemical resistance have made CFs a primary candidate for light-weight material applications. However, the high cost of CF production currently limits exploitation of these properties to military, aerospace and sporting industries[1,2]. As connected and automated vehicles (CAVs) have been reshaping the transportation sector, there is a growing desire to capitalize on this opportunity to enhance fuel efficiency and reduce emissions[3] by replacing metal components with light-weight, low-cost CF-reinforced composites. Over 90% of the current global CF market consists of polyacrylonitrile (PAN)-based CFs, with high melting point and superior carbon yield, which give rise to thermally stable and well-oriented CFs microstructures[4,5]. However, over 50% of their cost is attributed to the PAN precursor, a drawback that has triggered the search for low-cost alternatives.

To date, renewable and low-cost precursors such as pitch, lignin, rayon, and cellulose have been examined to reduce the cost of CFs, however, their poor mechanical properties impose usage limitations[6–9]. An alternative strategy for making CFs more cost effective in the automotive industry is the use of additive-reinforced composite CFs, in which the low-cost precursors are fortified by the addition of reinforcement components with superior mechanical properties such as carbon nanotubes (CNTs). For instance, Chae et al. showed that adding 1.0 wt.% CNT increases CF strength by 64% and modulus by 49%[10]. Papkov et al. also reported that addition of CNTs can decrease carbonization temperature and reduce both energy consumption and final production cost[11]. Addition of graphene oxide (GO) liquid crystals for CFs fabrication has been explored, however they exhibited low tensile strength due to their intrinsic defects, such as atomic defects of graphene sheets, nanoscale voids and boundaries, and randomly alignment of the graphene nanosheets [12–14]. More recently, we have shown the addition of a small amount of shear-exfoliated graphene (0.075 wt.%) in PAN polymer matrix can significantly improve the mechanical properties of CFs by decreasing the porosity of fibers. Inclusion of the graphene into the PAN matrix enhances Young's modulus to 233 GPa and the tensile strength to 1916 MPa, which depicts a 184% increase in modulus and a 225% increase in strength compared to PAN-based CFs without graphene inclusion[15]. However, these previous studies did not provide detailed information on the carbonization condition impacts or how the introduced additive affected the microstructure and final mechanical properties. A more detailed understating of how the carbonization conditions affects the CFs properties and how additives can change the local microstructures and the behind mechanisms at atomistic scale is essential to develop low-cost CFs based on alternative precursors. In this study, the impact of the carbonization temperature on PAN/graphene nanocomposite CFs are investigated by coupling atomic-scale simulation to experimental validation. Atomistic ReaxFF simulation and experimental results jointly unveil how graphene sheet acts as a seed for the growth of graphitic structure and modifies the arrangement of CF all-carbon membered rings under different carbonization temperatures. The addition of graphene enhances the alignment of 6-member carbon rings and minimizes pore size and defects which leads to superior mechanical properties. Compared to pure PAN-based CFs, the PAN/graphene composite CFs demonstrate higher strength and Young's modulus even at lower carbonization temperatures. As such, the demonstrated impacts of the graphene additives on PAN-based CFs could give us a pathway to decrease the final product cost of CFs by replacing expensive PAN precursor by low-cost graphene/polymer composite precursor and increasing energy efficiency by decreasing the carbonization temperature.

## 2. Simulation and Experimental Methods

### 2.1. Atomistic Scale ReaxFF Simulations

To achieve atomistic insights into PAN/graphene nanocomposite CFs and make comparison with current PAN-based CFs, reactive molecular dynamics (MD) simulations based on the ReaxFF potential are used. ReaxFF is an interatomic potential based on bond-orders between atoms and can be obtained by training the potential against quantum mechanics (QM) data, such as Density Functional Theory (DFT)[16,17]. In contrary to non-reactive MD simulations, ReaxFF potentials give us the capability to simulate bond formation/breakage, making it a powerful tool to investigate the chemistry of large systems, which are computationally too expensive to study by QM methods[17]. In this study, we implement the C/H/O/N-2019 force field parameters developed by Kowalik et al.[18] in order to simulate PAN/graphene CFs graphitic structure. Kowalik et al. reoptimized C/H/N-2010 forcefield parameters[19] with the focus on the stability of $N_2$ molecules. In this new forcefield, $N_2$ molecules are adequately stable, and thus do not react with carbon radicals, and frequent $N_2$ molecule removal during the carbonization simulations is not required[20]. In addition, the extension of C/H/N-2010 forcefield to C/H/N/O atoms paves the path to simulate a wide range of oxygen-containing (O-containing) polymer precursors such as oxidized PAN, PBO (poly(p-phenylene-2,6-benzobisoxazole)) and cellulose[18,21,22].

To build our systems, in order to simulate initial stages of carbonization process, a proposed oxidized PAN chain[20] and a hydrogen-terminated single-layer graphene sheet (16 Å × 16 Å ) are used after energy minimization **(Fig. S1.a, b)**. For both the PAN and PAN/graphene systems, 32 chains of oxidized PAN are randomly placed in the simulation box. In the PAN/graphene systems, a graphene sheet is placed in the center of the simulation box to replicate the graphene inclusion into the polymer matrix. The total numbers of atoms are 3648 and 3769 for the PAN and

PAN/graphene systems, respectively. The size of simulation cells is determined to obtain the initial density of precursors equal to 1.6 g/cm$^3$ for all the systems[18]. We apply periodic boundary condition in all directions.

Initially, NVT ensemble at 300K for 60ps is used to equilibrate the systems **(Fig. S1.c)**. In order to have statistics data for the initial stage of carbonization process, five system configurations within the last 5ps are heated up at a 10 K/ps rate to reach 2200K, 2500K and 2800K. Finally, five configurations for each temperature are utilized and kept at 2200K, 2500K and 2800K for 1 ns **(Fig. S1.d)**. The temperature difference between experiments and simulations is used to accelerate the reactions in simulations, therefore these can be sampled on the nanosecond time scale accessible by our MD simulations. A Berendsen-Anderson temperature thermostat with a damping constant of 100 fs is employed[23]. The time increment is 0.25 fs for each step of simulations. To unveil the graphene inclusion effects on PAN CFs, the catalytic role of graphene edges as seeds for graphitic structure growth is investigated. As N-containing groups play a significant role in graphitic structures evolution and converting carbon radical species into the graphitic network and O-containing groups contribute to initiating the carbonization[21], N/O-containing groups and gases are studied and various mechanisms for nitrogen gas production are proposed. The small gas molecule emission and carbon content of PAN/graphene CFs at different carbonization temperatures are evaluated. In addition, we have analyzed 5/6/7-membered all-carbon ring formations, which mainly construct the graphitic structure of CFs[24] along with the in-graphene and peripheral carbons. Ultimately, to further examine the graphitic structure of CFs, Radial Distribution Function (RDF) of carbon-carbon bonds and alignment of the 6-membered all-carbon rings are investigated. The alignments of the 6-membered all-carbon rings are calculated based on

Herman's Orientation Function (HOF). HOF is defined as $HOF = \frac{1}{2}[3\langle cos^2\theta \rangle] - 1$ where $\theta$ is the orientation angle of the 6-membered all-carbon rings relative to the axis of interest[25].

## 2.2 Synthesis of PAN and PAN/Graphene Carbon Fibers

Pure PAN and PAN with 0.075 wt.% graphene (PAN/graphene) precursor fibers were produced by a lab-scale wet-spinning setup, in which PAN (Goodfellow, mean particle size: 50 μm, copolymer: 99.5% AN/0.5% MA, molecular weight: 230 kg /mol) was used as precursor powder, DMSO as the solvent and the coagulation bath is DMSO/DI-$H_2O$ system. The added graphene was prepared through a modified shear-exfoliation method. The details of graphene preparation, PAN/DMSO spinning dope, and spinning parameters can be found in our previous work[15]. Briefly, the 10 wt.% PAN or PAN/graphene/DMSO solution was first prepared and then extruded through a 5 mL syringe with a needle diameter of 110 μm. The extruded fiber went through two consecutive coagulation baths (DMSO/DI-$H_2O$ = 7:3 and 3:7 by volume, respectively) followed by a washing bath. The obtained precursor fibers were dried at 80 °C for 12 h, oxidized at 250 °C for 2 h in air and then carbonized at three different temperatures, of 1000, 1250, and 1500 °C in Ar, holding for 0.5 h, respectively.

## 2.3 Experimental Characterization of CFs

Field-emission scanning electron microscopy (FE-SEM, FEI Quanta 650) was used to investigate the morphology of CFs. X-ray diffraction (XRD) and Raman patterns were recorded by Pro Multi-Purpose diffractometer (MPD) equipped with Cu Kα radiation with λ = 0.15406 nm in a 2$\theta$ range from 10° to 80° and Raman spectroscopy (Renishaw InVia equipped with a 514 nm laser beam), respectively. Tensile tests were carried out by an MTS Nano Bionix® testing system with a 0.5 N load cell and a displacement rate of $10^{-4}$ mm $min^{-1}$. The gauge length of the samples was fixed at

20 mm in this work. Young's modulus, $E$, is calculated based on the stress-strain curve in the elastic region.

## 3. Results and Discussion

### 3.1. Atomistic-Scale Modeling of the PAN/Graphene Nanocomposite CFs

In order to acquire atomistic perspective on how the graphene inclusion affects chemistry, structure and mechanical characteristics of PAN/graphene CFs in the initial stage of carbonization process, ReaxFF molecular dynamics (MD) simulation are employed. A detailed investigation of the MD simulation trajectories unveil that graphene surface carbon atoms are relatively inactive and do not form bonds with the polymer matrix even at high temperatures, due to their conjugated $sp^2$ electronic configuration. On the other hand, the graphene edges have catalytic effects on the formation and evolution of graphitic structures as illustrated in **Fig. 1**. At the beginning of the carbonization process, the graphene sheet is intact and there are no chemical bonds between the graphene sheet and polymer matrix. The graphene sheet is hydrogen terminated in order to hinder low-temperature reactivity of graphene edges **(Fig. 1a)**. As we anneal the system, C-H bonds at the graphene edges break and the hydrogen atoms migrate from the graphene sheet. The undercoordinated carbon atoms at the graphene edges form bonds with carbon or nitrogen atoms which belong to PAN chains. The C-C and C-N bond formations at the graphene edges lead to the transformation of the PAN molecule to heteroatoms ring structures including 5, 6 or 7-membered rings **(Figs. 1b-d)**. These N-containing rings eventually release their nitrogen atoms and become all-carbon membered rings **(Figs. 1e-j)**. For instance, a 5-membered N-containing ring, presented in **Fig. 1d** by a green circle, is formed next to two 7-membered N-containing rings (red and orange circles in **Fig. 1d**) at the edge of graphene. The C-N bond in the 7-membered ring breaks, and the 7-membered N-containing ring alters to a 6-membered all-carbon ring. Eventually, this 6-

membered all-carbon ring transforms to a 7-membered all-carbon ring (red circles in **Figs. 1d-j**). Simultaneously, the other 7-membered N-containing ring changes to a 6-membered all-carbon ring by losing its nitrogen atom (orange circles in **Figs. 1d-j**). At the same time, the C-N bonds in the 5-membered N-containing ring dissociate and are replaced with a C-C bond to form a 5-membered all-carbon ring (green circles in **Figs. 1d-j**). The graphitic structure expands by forming more all-carbon membered rings until the end of the simulation at 1 ns **(Figs. 1k-l)**. The limited number of observed all-carbon membered rings formed at the edge of graphene is due to the simulation time scale. Since we perform only 1 ns MD simulations, compared to hours of carbonization in experiments, only a few all-carbon membered rings are observed in our simulations. This indicates that we can only simulate the initial stages of carbonization process. However, these MD simulations still provide valuable insights and are able to uncover the catalytic role of graphene edges as seeds in graphitic structure development.

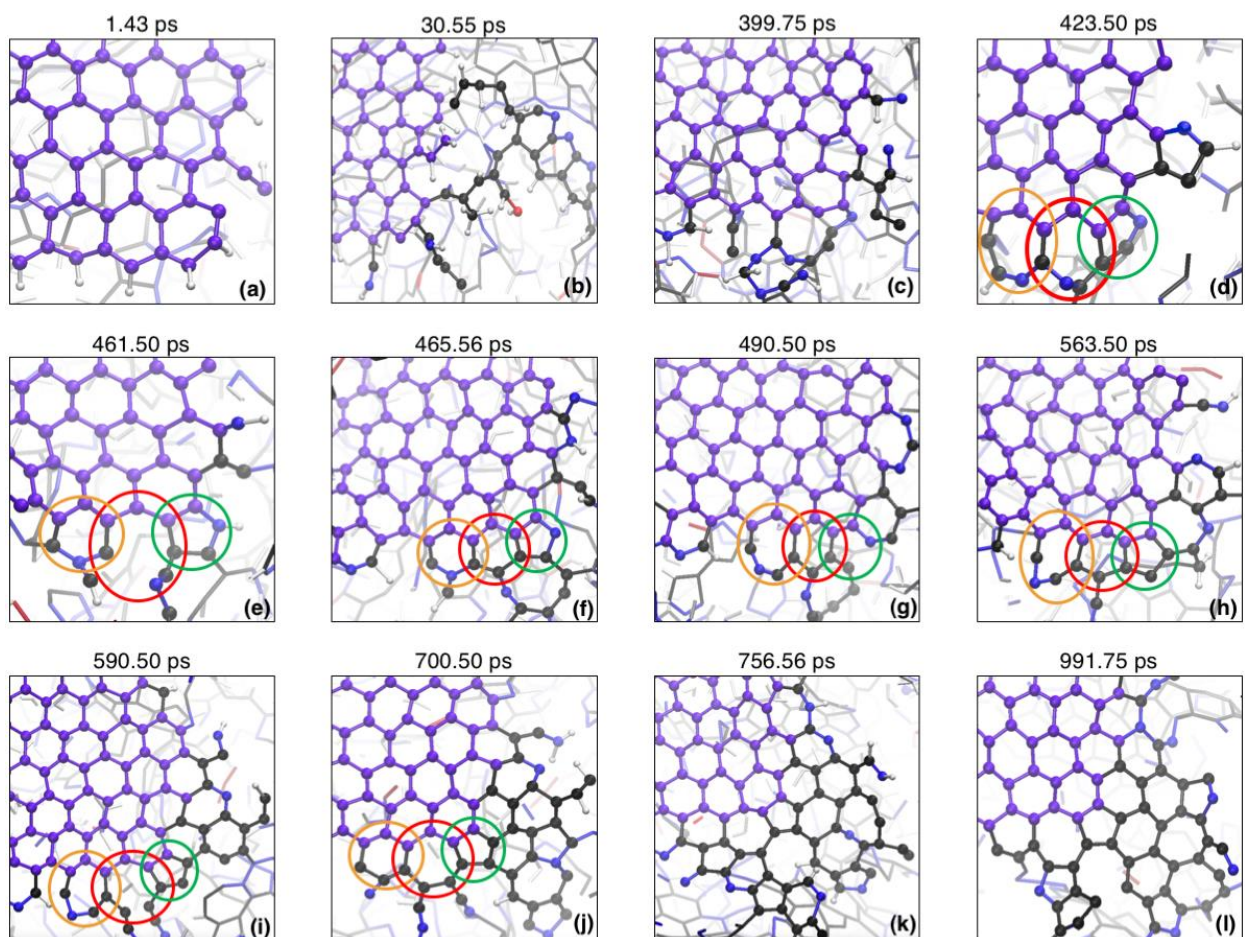

**Fig. 1. Atomistic ReaxFF simulations of the initial stage of carbonization process for PAN/graphene precursor.** Snapshots of the PAN/graphene CFs during the carbonization process to show the catalytic role of graphene edges as seeds in graphitic structure growth, while the graphene surface carbon atoms are relatively inactive due to their conjugated $sp^2$ electronic configuration. Formation and evolution of 5/6/7-membered all-carbon rings take place at the graphene edges. (a) Intact graphene sheet at the beginning of the carbonization process. (b-d) carbon atoms at the graphene edges transform the PAN molecule to heteroatoms ring structures by forming bonds with carbon or nitrogen atoms of the PAN molecule. (e-j) N-containing rings release their nitrogen atoms and become all-carbon membered rings. Color code circles are utilized to aid tracking the rings. (k, l) the graphitic structure expands by forming more all-carbon membered rings. Purple spheres represent initial graphene structure. Carbon, nitrogen, oxygen, and hydrogen atoms are represented in black, blue, red, and white. These snapshots were saved every 6.25 fs of the PAN/graphene carbonization process at 2500K.

To better understand the pathways of all-carbon membered ring formations, the in-graphene and peripheral carbons (**Fig. 2g**) out of the all-carbon ring networks are analyzed (**Figs. 2a and d**). The in-graphene carbon is defined as a carbon atom only having 3 carbon neighbors ($sp^2$ carbon) that all belong to all-carbon ring (3- to 8-membered) networks; and the peripheral carbon is defined as a carbon atom connecting to 1 or 2 carbon neighbors that belong to the all-carbon ring networks, corresponding to the edge growth of the graphitic ring structures. We count the in-graphene and peripheral carbon atom conversions of PAN/graphene CFs in various carbonization temperatures, while the initial graphene carbons are not included. For the in-graphene carbon conversion, a direct proportionality with the carbonization temperature is observed. Higher carbonization temperatures lead to higher in-graphene carbon conversions and the carbon conversions keep growing in the entire carbonization process, which indicates the continuous tendency of the systems to form in-graphene carbons. Interestingly, at the beginning of the carbonization process (up to 250 ps) the conversion rate of in-graphene carbons at 2800 K is higher than 2500 K and 2200 K. However, the conversion rates of in-graphene carbons are almost equal after the first 250 ps. On the other hand, after 1 ns carbonization, the peripheral carbon conversions at different carbonization temperatures converge to a plateau, while the in-graphene carbon conversions still have a positive slope. This indicates that by continuing the simulation, the peripheral carbon conversions remain constant, while the in-graphene carbon conversions increase, and larger graphitic structure can be achieved. The peripheral carbon growth at 2800 K appears to have the sharpest slope at the beginning of the carbonization, but it converges to a plateau at around 250 ps. This suggests that the edges of the all-carbon membered ring networks quickly become stable at 2800 K. The peripheral carbons at 2500 K have the second sharpest slope in the beginning, but after 500 ps it exceeds those carbonized at 2800 K. This is because the functional groups at 2800 K are prone to quickly having the ring openings and all-carbon membered ring edge formations, but the functional groups at 2500 K tend to gradually participate in the graphene edge formations. This trend can also be observed in PAN-based CFs **(Fig. S2)**.

MD simulation trajectories suggest that the N-containing groups and gases in PAN/graphene CFs play an important role in the evolution of graphitic structure by converting carbon radical species into the graphitic network. This agrees with our previous findings on PAN/PBO blends CFs[21].

Therefore, we analyze the role of the N and O-containing groups/gasses for PAN/graphene CFs and how the carbonization temperature affects them. We also propose various mechanisms for $N_2$ production, as one of the main gases, emitted during the carbonization process. The evolutions of N-containing and O-containing species at different carbonization temperatures are shown in **Figs. 2b,c,e,f**. The N-containing groups consist of the amine groups, imine groups, pyridine-like groups and nitrile groups, and the N-nitrogen containing gases are mainly $NH_3$ and $N_2$. The O-containing groups include the carbonyl groups, hydroxyl groups and the bridging ether links, and the O-containing gases are mainly $H_2O$, $CO_2$ and CO. The details about the functionality of the N and O-containing species can be found in our previous work[21]. We normalize the N-containing species and O-containing species to a scale from 0 % to 100 %, based on the total numbers of nitrogen and oxygen atoms originally in the systems. The production of N-containing and O-containing gases result in a reduction of N-containing and O-containing groups in all three carbonization temperatures (**Figs. 2b,c,e,f**). Higher carbonization temperatures result in higher N and O-containing gases, which can lead to CFs with higher carbon contents. The increase in N and O-containing gases correspond to the depletion in N and O-containing functional groups. In comparison with N-containing groups, O-containing groups decay rapidly in the early stage of the carbonization simulations, quickly converting to O-containing gases. The N-containing groups, however, slowly evolve into the N-containing radicals, which are still reactive. These reactive N-containing groups reside inside the polymer molecules and carbon ring networks which would assist the further carbonization (**Fig. 1**). Moreover, a lower temperature yields a lower rate of conversion for both N-containing and O-containing species. When the temperature rises to 2800 K, a significant turning point appears at around 100 ps for O-containing species, leading to much slower slope for the reaction until reaching a plateau. A similar behavior is found at around 400 ps for N-containing species at 2800 K. We find that the PAN-based CFs exhibit the same behavioral evolution (**Fig. S2**).

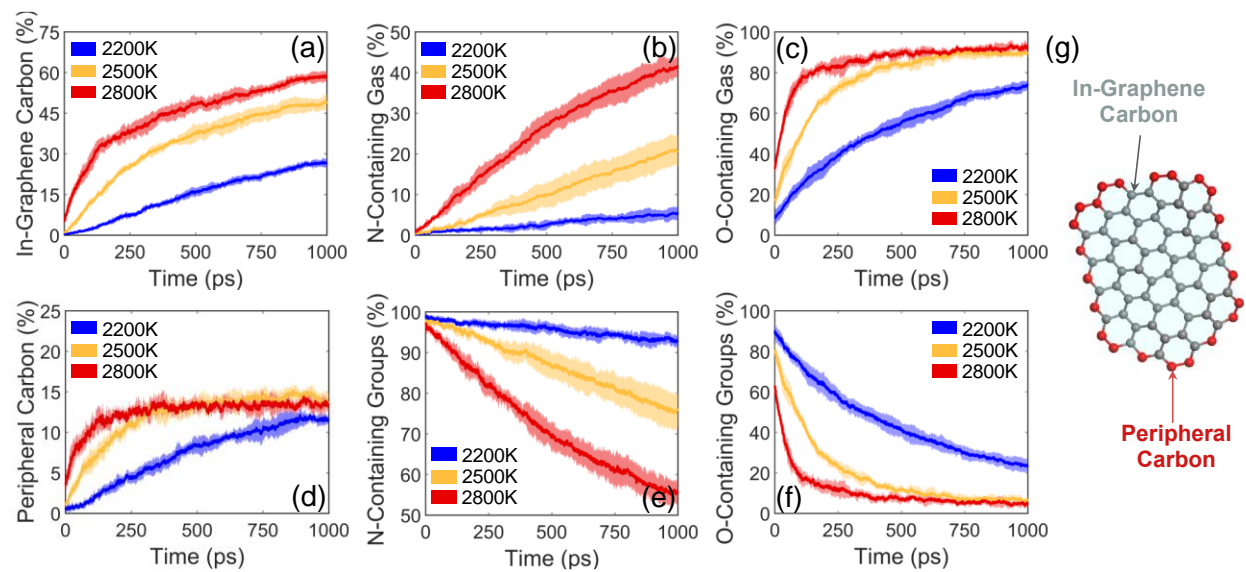

**Fig. 2. Classification of the carbon atoms out of the graphitic network and evolution of N and O-containing species at different carbonization temperatures.** (a) In-graphene carbons, (b) Peripheral carbons, (c) N-containing gases, (d) N-containing groups, (e) O-containing gases, (f) O-containing groups, (g) The schematic of in-graphene carbons (gray spheres) and peripheral carbons (red spheres). The N-containing groups consist of the amine groups, imine groups, pyridine-like groups and nitrile groups, and the N-containing gases are mainly $NH_3$ and $N_2$. The O-containing groups include the carbonyl groups, hydroxyl groups and the bridging ether links, and the O-containing gases are mainly $H_2O$, $CO_2$ and CO. nitrogen-containing groups consist of the amine groups, imine groups, pyridine-like groups and nitrile groups, and the nitrogen- containing gases are mainly $NH_3$ and $N_2$. The curves are plotted using the averaged data by 5 samples (dark colors), and the standard deviations are presented by the transparent shadows.

We propose various $N_2$ production mechanisms in PAN/graphene CFs based on the MD trajectories. Graphene inclusion into PAN matrix offers a new pathway for $N_2$ production in which graphene edges act as a host for nitrogen atoms (**Fig. 3**). As illustrated in **Fig. 3**, at the beginning of the MD simulation the PAN chain containing nitrogen atoms is intact (**Fig. 3a**). By annealing the PAN/graphene system to mimic the initial stage of carbonization, a carbon chain containing the nitrogen atom, N1, is detached from the PAN chain, and it migrates toward the graphene sheet and makes a bond with an undercoordinated carbon atom at the graphene edge (**Figs. 3b-d**). The attached carbon chain transforms to a N-containing ring (**Fig. 3e**). The same scenario happens to the second nitrogen atom, N2 **(Figs. 3f-i).** These two adjacent N-containing rings open and transform to a ring at the graphene edge in which N1 and N2 are bonded **(Figs. 3j-o)**. These two bonded nitrogen atoms leave the graphitic structure as an $N_2$ molecule. Although extensive studies have been done on selective edge-functionalized graphene for energy conversion and storage[26,27], dye-sensitized solar cells[28], sensors[29] and enhanced oxygen evolution reaction (OER) [30], to the best of our knowledge there has not been any study on adding edge-functionalized graphene to PAN-based CFs. The edge-functionalized graphene could provide a good host for the formation of peripheral N-containing groups constantly as the mobile N-containing polymer residues move nearby, where the N-containing groups are efficient for capturing and converting the carbon radical species into expanded graphitic networks. In the meantime, the N-containing groups evolve into volatile gases, thus the graphitic structure could continuously grow. This mechanism aligns well with our previous study [21]. On the other hand, the graphitic-N groups (also called quaternary nitrogen groups), the nitrogen atoms bonded with 3 neighbor carbon atoms in the basal plane of graphene, are far less reactive than the N-containing groups on the edges of graphene, and they could hamper the graphitic structure formation[31–33]. This mechanism suggests that adding selective edge-functionalized graphene to polymer matrix instead of free-standing graphene could serve as a catalyst to capture the carbon radical species and eliminate more nitrogen and to enhance the carbon content and eventually the quality of CFs. However, with the current simulation setup, carbonization of CFs with nitrogen edge-functionalized graphene nanocomposite is not feasible, since we are limited to high temperature range (above 2200K) to be able to observe formation of all-carbon membered rings. At the high temperature range, the nitrogen atoms dissociate from the graphene sheet and the graphene sheet behaves similarly to pristine graphene. In the second scenario, $N_2$ formation involves reaction between a PAN chain and an ammonia molecule. The C-

N bond in a pyridine-like 6-membered ring on a PAN chain is broken and the nitrogen atom reacts with the ammonia to form $N_2$ **(Fig. S3)**. Thirdly, nitrogen is formed by the reaction between two ammonia molecules, in which each ammonia loses one hydrogen atom and the resultant $NH_2$ radicals react and form a hydrazine molecule. The N-H bonds in hydrazine break and $N_2$ is eliminated **(Fig. S4)**. These three reaction mechanisms have not been considered previously. In addition to the aforementioned mechanisms, two other well-known mechanisms in PAN-based CFs, intermolecular and intramolecular are also observed[34]. In the intramolecular mechanism, the C-N bonds of two neighbor pyridine-like 6-member rings on one PAN chain break and eventually $N_2$ is formed. **(Fig. S5)**. However, the intermolecular mechanism involves the reaction between two nitrogen atoms on two different PAN chains **(Fig. S6)**.

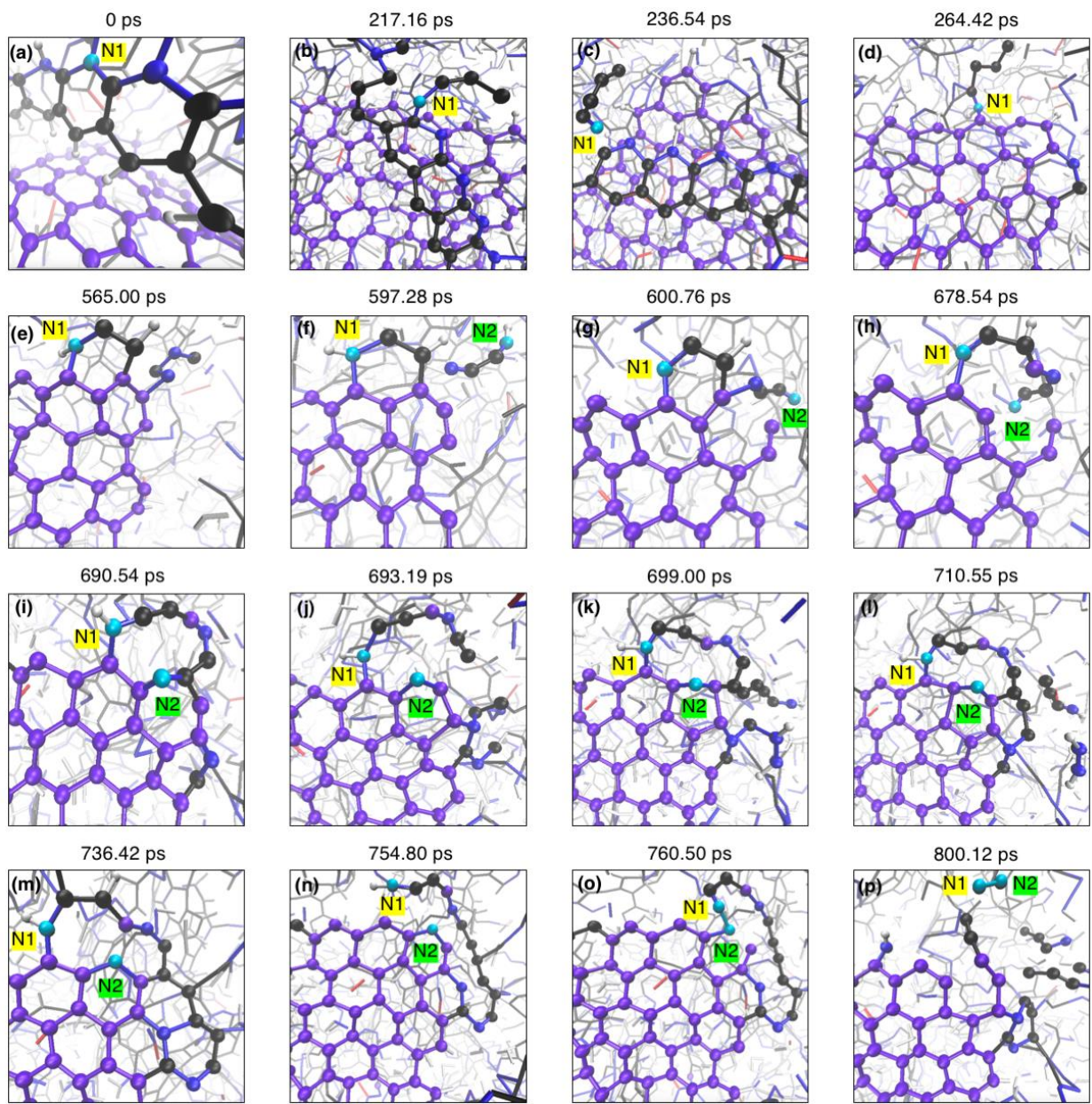

**Fig. 3. Graphene edges introduce a new pathway for $N_2$ production.** Two carbon chains containing nitrogen atoms from different PAN chains, migrate from the PAN chains, bond with undercoordinated carbon atoms at graphene edges and form nitrogen containing rings. The transformations of these N-containing rings lead to formation of $N_2$ molecules. Purple spheres represent the initial graphene structure. Carbon, nitrogen, oxygen, and hydrogen atoms are represented in black, blue, red, and white. We colored the nitrogen atoms involved in $N_2$ formation as cyan and labeled them as N1 and N2. These snapshots were saved every 6.25fs of the carbonization process.

Based on the gas production analyses, the higher carbonization temperatures give rise to higher small molecule gas productions such as $N_2$, $H_2$ and $H_2O$ which are the major gas emissions during the carbonization process (**Figs. 4a-c**). The higher amount of gas production in the CFs leads to the higher carbon content of PAN/graphene CFs. At each carbonization temperature, the PAN/graphene CFs shows slightly higher carbon contents compare to the PAN CFs **(Fig. 4d)**.

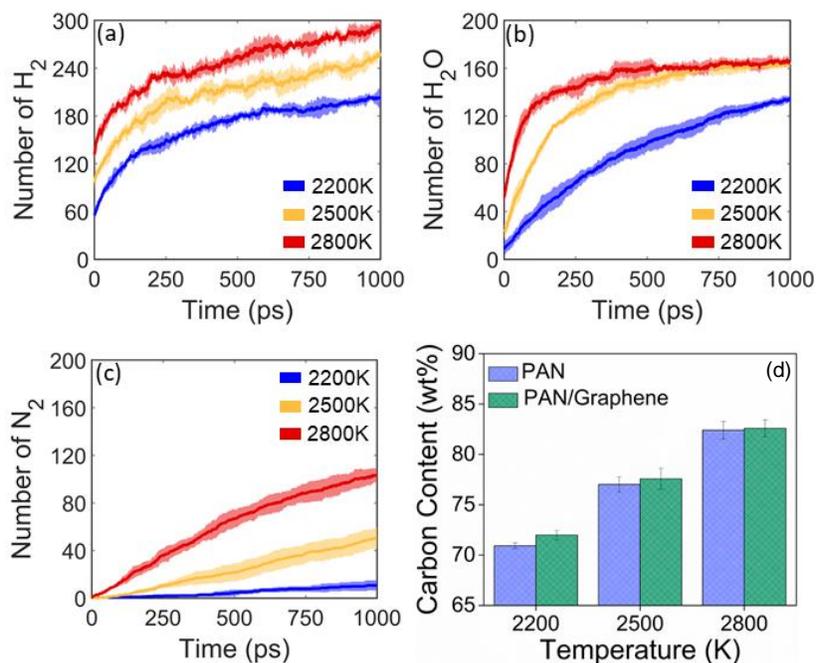

**Fig. 4. Gas emission analyses of PAN/graphene nanocomposite CFs during carbonization and carbon contents of two fibers.** Productions of (a) $H_2$ (b) $H_2O$ (c) $N_2$. (d) Carbon contents of PAN/graphene nanocomposite CFs at various carbonization temperatures.

Since CFs mainly consists of all-carbon rings, a quantitative analysis of the 5/6/7-membered all-carbon rings is highly relevant. Zhu et al. have implemented the C/H/O/N-2019 force field parameters and illustrated that for PAN-based CFs, no significant all-membered carbon rings are formed at 1800K[35] under MD-accessible time-scales. In order to overcome these limitations and be able to observe all-membered carbon ring formation at nanosecond scale, we perform the simulations at temperatures higher than 1800K. For PAN/graphene CFs, all-carbon membered ring formation can be seen at 2200 K, 2500K and 2800 K. Higher carbonization temperature leads to increase in the formation of 5/6/7-membered all-carbon rings. At each carbonization temperature, more 6-membered all-carbon rings are formed compared to 5 and 7-membered all-carbon rings corresponding to the topological defects[36], which indicates the tendency of the systems to form 6-membered all-carbon rings.

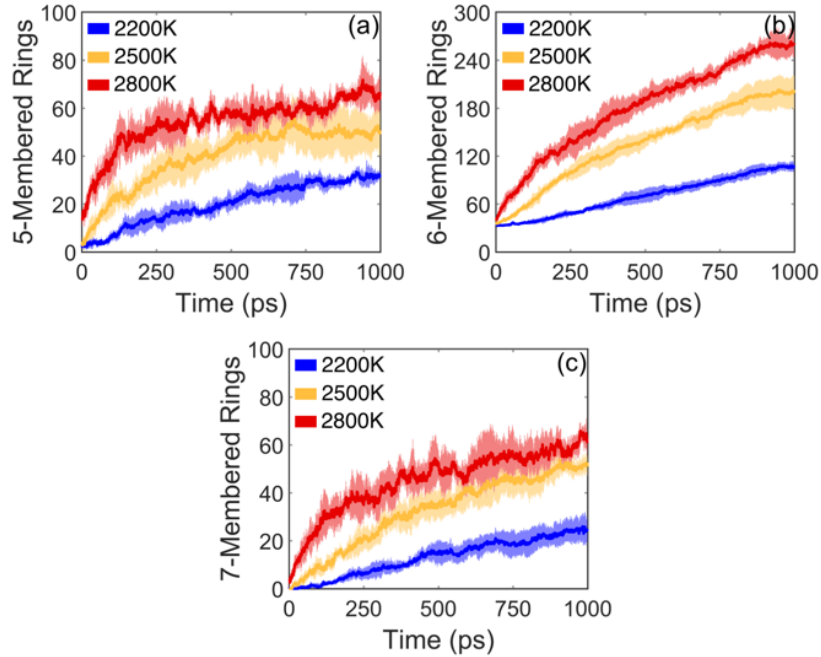

**Fig. 5. All-membered-carbon rings formation of PAN/graphene nanocomposite CFs at various carbonization temperatures.** (a) 5-membered all-carbon membered rings, (b) 6-membered all-carbon membered rings (6-membered all-carbon membered rings of the initial graphene sheet are included), (c) 7-membered all-carbon membered rings.

In addition to the number of all-membered carbon rings, the alignment of these rings significantly affects the mechanical properties of CFs by regulating the microstructure of CFs[15]. Here we employ HOF to calculate the alignment of 6-membered all-carbon rings. At each carbonization temperature, graphene inclusion increases the alignment of 6-membered all-carbon rings of the system remarkably which leads to the microstructure enhancement. Although PAN/graphene CFs at 2200K exhibits the highest alignment percentage, it is not expected that PAN/graphene at 2200K shows the best mechanical properties. It is crucial to consider the number of all-membered carbon rings in predicting the performance of the CFs, and at 2200K we have the least number of all-membered carbon rings. The all-membered carbon ring formation could provide the nucleation site for more extensive graphitic structure growth, and structurally the growth can be characterized by the partial carbon-carbon (C-C) radial distribution function (RDF), as shown in **Fig. 6**. Significant increases in peak intensities that correspond to the C-C 1st, 2nd, and 3rd neighbors can be observed as the carbonization temperature increases from 2200 K to 2800 K. The 1st, 2nd, and

3rd peaks associate with the distances between a target carbon atom and its 1st, 2nd, and 3rd $sp^2$ bonded neighbors. The higher intensity at the designated distances represent that a higher proportion of carbon are graphitized. The peaks at the 1st neighbor distance of 2500 K and 2800K exhibit roughly identical intensities, but that at 2800K demonstrates slightly higher intensities at the 2nd and 3rd C-C neighbor distances (**Fig. 6b**). This can be attributed to the significant $sp^2$ bonded amorphous carbon growth at the edges of the embedded single layer graphene, and the carbon atoms at 2800K should tend to be better carbonized if the carbonization dwell time is prolonged. The massive edge growth at 2500 K can be also seen from **Fig. 2b**, where the carbon atoms on the edges of the graphene start becoming $sp^2$ bonded, but not forming perfect graphitic rings. Unlike the PAN with graphene, the PAN systems exhibit distinct peak intensities at the given 3 temperatures, as shown in **Fig. S7**, suggesting that the further carbonization initiated on the edges of the embedded graphene is associated with the formation of the $sp^2$ carbons.

Based on the ReaxFF MD simulation results, we expect superior mechanical properties of PAN/graphene CFs compare to PAN CFs at each carbonization temperatures due to improved alignments of 6-member all-carbon rings which leads to minimized pores and defects. In addition, enhanced PAN/graphene CFs mechanical properties with increase in carbonization temperatures is expected due to higher amount of all-carbon membered rings formation and carbon content. Later in this paper, these results will be verified by experimental mechanical testing.

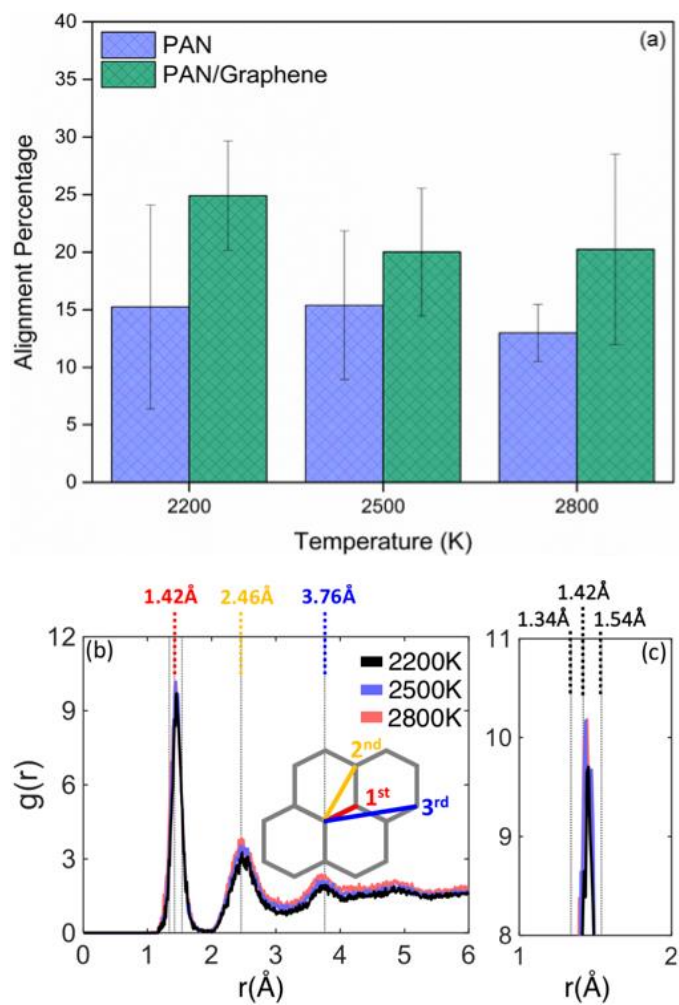

**Fig. 6. Alignment percentage and radial distribution function (RDF) of PAN/graphene nanocomposite CFs at various carbonization temperatures.** (a) Alignment percentage of PAN/graphene nanocomposite CFs based on HOF calculations, (b) Carbon-carbon (C-C) RDF, denoted as g(r) in arbitrary unit for intensity, for PAN/graphene at temperatures of 2200 K, 2500 K, and 2800 K. The red, yellow, and green dash lines and values signify the distances r(Å) between the target carbon atom and its 1st, 2nd, and 3rd $sp^2$ bonded neighbor carbon atoms, respectively. The dash lines from left to right on (c) correspond to the bond lengths of double bonded, $sp^2$ bonded, and single bonded C-C bonds, which are equal to 1.34 Å, 1.42 Å, and 1.54 Å, respectively. Each RDF curve is plotted using the averaged data by 5 samples.

**3.2 Experimental Characterization of PAN and PAN/graphene Nanocomposite CFs Carbonized at Different Temperatures**

PAN and PAN/graphene nanocomposite CFs at different carbonization temperatures were produced and analyzed to verify the atomistic scale simulation results. In this work, the effects of carbonization temperature on structure and mechanical properties of PAN/graphene CFs at different carbonization temperatures have been investigated. Here, we chose the representative PAN and PAN/graphene CFs, carbonized at 1000 ºC, 1250 ºC and 1500 ºC to investigate the carbonization temperature effect on the fibers' structural and mechanical properties. The XRD patterns of the carbonized PAN and PAN/graphene CFs produced at different carbonization temperatures are shown in **Figs. 7a and b**. A weak and broad peak can be detected at around 24º for both carbonized PAN and PAN/graphene CFs at different carbonization temperatures, which can be indexed to the (002) plane of the graphitic structure. The similar diffraction peaks for both PAN and PAN/graphene indicate the carbonization starting at a low temperature and increasing the temperature from 1000 ºC to 1500 ºC does not change the fibers' crystal structure significantly. The reason why there is no notable change in the XRD patterns for both PAN and PAN/graphene is the relative low carbonization temperature. Usually, the temperature needs to be increased more than 2000 ºC in order to see the clear revolution in graphitic structure[35]. In addition, the Raman spectra of the PAN and PAN/graphene nanocomposite CFs are compared in Figure 7, which can be used to characterize the ordered/disordered structure of graphitic materials (**Figs. 7c and d**). The typical D band at 1350 cm$^{-1}$ is ascribed to the breathing mode of point photons of $A_{1g}$ symmetry, while the G band at ~1585 cm$^{-1}$ results from the first-order scattering of the $E_{2g}$ phonon mode of in-plane sp$^2$ C atoms[35]. The intensity ratio of D and G peak, denoted as $I_D/I_G$, can be used to measure the quantity of ordered graphitic structures[37,38]. The comparison of the $I_D/I_G$ ratio of the

PAN and PAN/graphene CFs carbonized at different carbonization temperatures are shown in Table 1. Overall, the $I_D/I_G$ ratio for both PAN and PAN/graphene CFs decreased with increasing carbonization temperature, indicating formation of more ordered graphitic structure. At the carbonization temperature of 1500 °C, the $I_D/I_G$ ratio of PAN/graphene CF is much smaller than that of PAN CF, which suggests that the addition of graphene leads to formation of more graphitic structure.

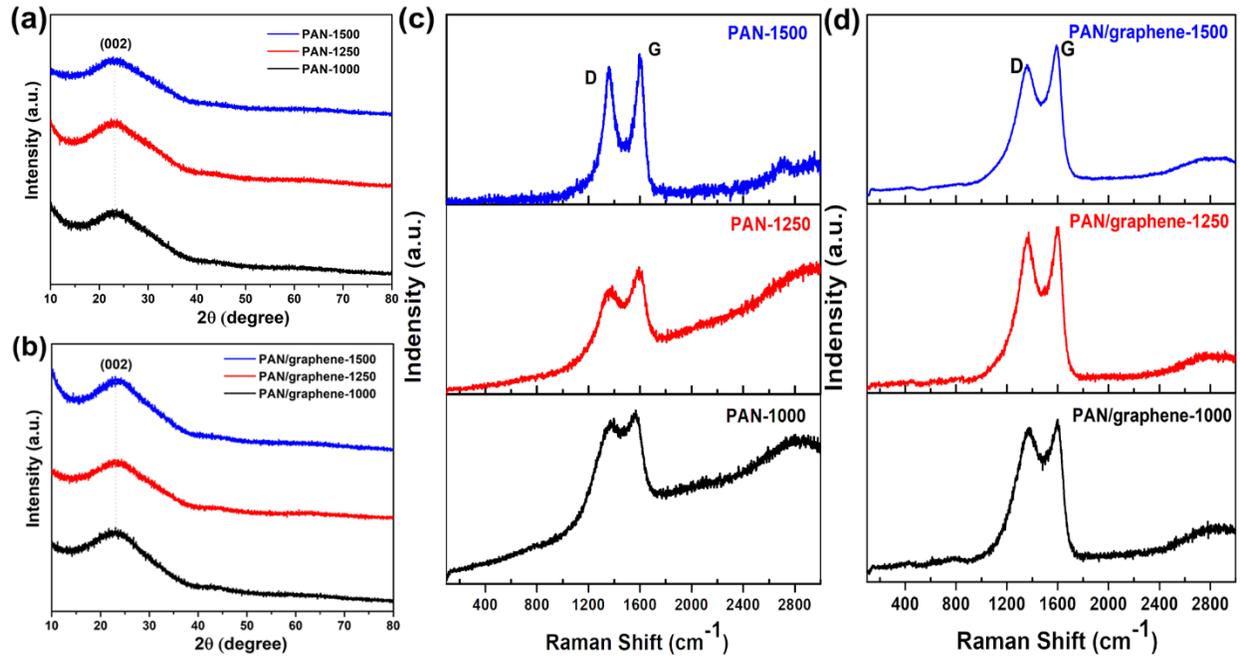

**Figure 7. X-ray diffraction (XRD) and Raman spectra of PAN CFs and PAN/graphene nanocomposite CFs.** XRD patterns of the carbonized PAN (**a**) and PAN/graphene (**b**) CFs at different temperatures; Raman Spectra of the carbonized PAN (**c**) and PAN/graphene composite (**d**) CFs at different temperatures.

**Table 1.** Comparison of the $I_D/I_G$ ratio obtained from the Raman spectra of PAN and PAN/graphene at different carbonization temperatures.

| Sample | 1000 °C | 1250 °C | 1500 °C |
|---|---|---|---|
| **PAN** | 0.945 | 0.897 | 0.901 |
| **PAN/graphene** | 0.984 | 0.943 | 0.886 |

The cross-sectional SEM images of the carbonized PAN (**Figs. 8a-c**) and PAN/graphene (**Figs. 8d-f**) fibers at different carbonization temperatures reveal that the addition of graphene into the PAN could significantly change the fibers' microstructure. Compared with PAN CFs, the PAN/graphene CFs demonstrate minimized pore size and much denser cross-sectional surface, which is essential for high-performance CFs. The detailed microstructure-modification mechanism of graphene has been discussed in our previous work[15], and can be ascribed to the favorable edge chemistry and improved PAN polymer chain alignment because of the appearance of graphene. Clearly, both PAN and PAN/graphene CFs keep their original microstructure with the temperature increase from 1000 °C to 1500 °C, which suggests that carbonization temperature does not change the overall morphology of PAN or PAN/graphene CFs in this temperature range. Compared to pure PAN CFs, PAN/graphene CF demonstrate much increased tensile strength (**Fig. 8g**), Young's modulus (**Fig. 8h**) and fracture strain (**Fig. 8i**) with increasing carbonization temperature. Specifically, for PAN/graphene CFs, when the carbonization increased from 1000 °C to 1500 °C, the strength increases from 668 to 1922 MPa (**Fig. 8g**) and the Youngs' modulus increases from 105 to 253 GPa (**Fig. 8h**), which represents a 187.7% increase in strength and a 140.9% increase in modulus, indicating that the increased carbonization temperature significantly improves the fibers' mechanical properties. The increased mechanical properties of PAN/graphene CF compare to pure PAN CFs can be ascribed to the reduced porosity, increased polymer chain alignment in the microstructure (**Figs. 8d-f**), and the increased carbon ring-alignment of 6-membered all-carbon rings (**Fig. 6**) at elevated temperatures, which are caused by the introduction of graphene. However, we did not see a significant increase in terms of the tensile modulus at the carbonization temperature of 1500 °C for pure PAN CFs, which can result from the porous microstructure of the lab synthesized PAN precursor, the higher carbonization temperature introduced more structural defects, like the cavity and boundaries, which will cause the stress concentration for PAN CFs. Even though, when we compared the mechanical properties of PAN CFs and PAN/graphene CFs, at the same carbonization temperature, the overall mechanical properties of PAN/graphene CFs are much better than those of the PAN CFs, which suggests the importance of graphene addition. In addition, the PAN/graphene CFs carbonized at 1250 ºC demonstrate 90.9% increase in strength (from 632 to 1207 MPa and 101.9% enhancement in Young's modulus (from 88 to 177 GPa) compare to the pure PAN CFs carbonized at 1500 ºC. The superior mechanical properties of PAN/graphene CFs at lower carbonization temperatures offer a pathway to both energy savings and cost reduction by decreasing the carbonization temperature and could provide key insights for the development of low-cost CFs.

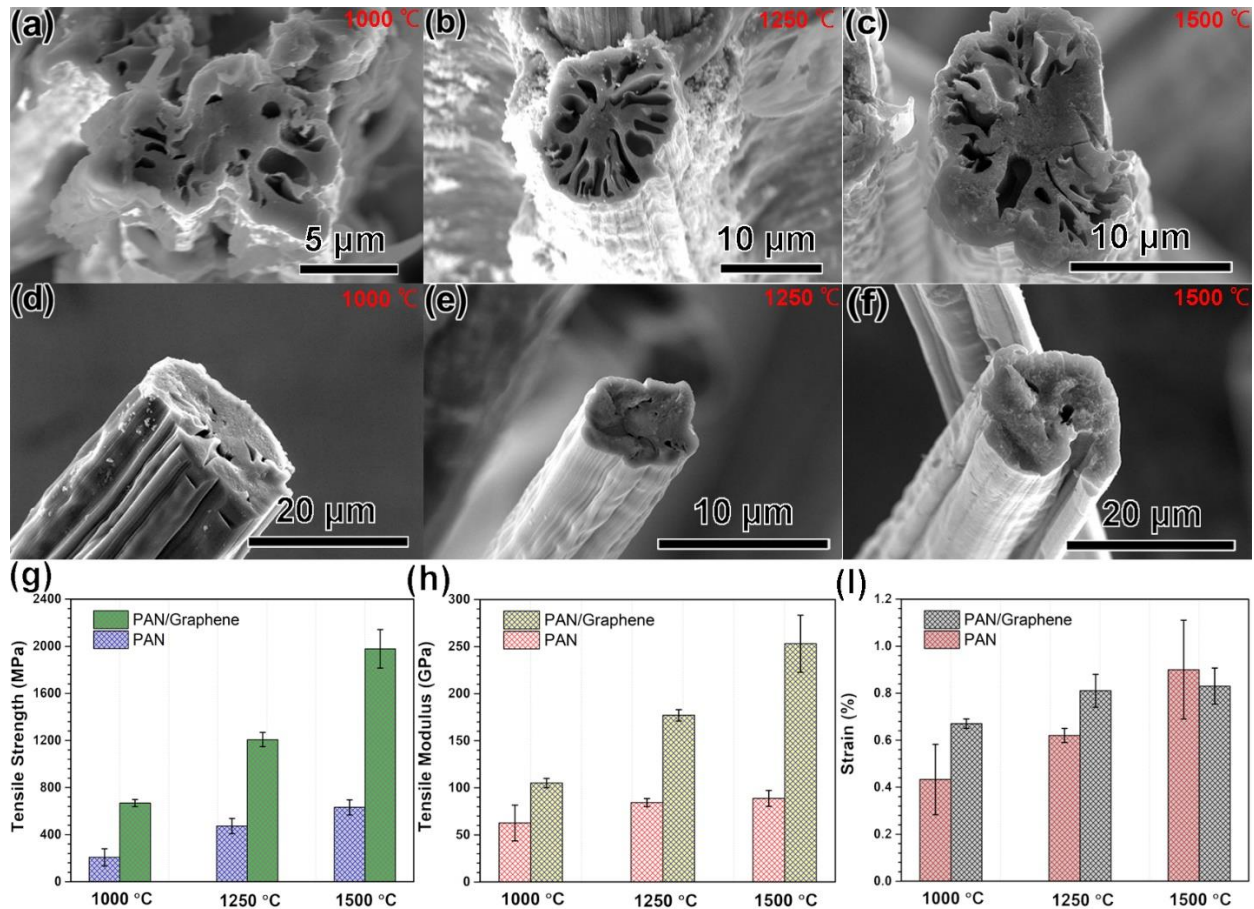

**Fig. 8. Cross-sectional SEM images and mechanical properties of PAN CFs and PAN/graphene nanocomposite CFs**. Cross-sectional SEM images of the carbonized PAN CFs at different temperatures, (**a**) 1000 °C, (**b**) 1250 °C (**c**) 1500 °C; Cross-sectional SEM images of the carbonized PAN/graphene composite CFs at different temperatures, (**d**) 1000 °C, (**e**) 1250 °C (**f**) 1500 °C; Tensile strength (**g**), Young's Modulus (**h**), and fracture strain (**i**) of carbonized PAN and PAN/graphene composite CFs.

## 4. Conclusions

In this study, atomistic scale simulations and experimental characterizations are jointly employed to uncover carbonization condition impacts on PAN/graphene CFs properties and how the addition of graphene affects the carbon rings formation, alignment of 6-membered all-carbon rings, as well as the mechanical properties of PAN/graphene CFs at different carbonization temperatures. Our simulations demonstrate that the graphene edges have a catalytic role and act as seeds for the growth of graphitic structure with improved alignment. The graphene sheet can guide the formation of nitrogen-containing groups as a key component in capturing and converting carbon radical species into graphitic network. We find that a higher carbonization temperature leads to more 6-membered all-carbon rings, higher gas emissions and increased carbon content of the PAN/graphene nanocomposite CFs. Raman spectroscopy inspection demonstrating reduced $I_D/I_G$ ratio validates the simulation results, indicating higher content of graphitic structure resulting from increased carbonization temperature. The PAN/graphene CFs exhibit denser microstructure, better alignment of 6-membered all-carbon rings, and higher content of graphitic structure compared to the PAN CFs at elevated carbonization temperatures, which leads to superior mechanical properties over that of PAN CFs. Interestingly, mechanical characterizations indicate that PAN/graphene CFs carbonized at 1250 ºC demonstrate 90.9% increase in strength and 101.9% enhancement in Young's modulus compare to the PAN-based CFs carbonized at 1500 ºC, which indicates that addition of graphene provides an efficient way to energy saving and cost reduction of CFs by lowering the carbonization temperature while retaining desirable mechanical properties.

**Acknowledgements**

We gratefully acknowledge the support from the U.S. Department of Energy (DOE), Vehicle Technologies office, under contract number DE-EE0008195.

# Supporting Information

# Polyacrylonitrile (PAN)/Graphene Nanocomposite: Toward Next Generation Carbon Fibers

*Siavash Rajabpour[1], Qian Mao[2], Zan Gao[3], Mahdi Khajeh Talkhoncheh[1], Jiadeng Zhu[3], Yosyp Schwab[3], Malgorzata Kowalik[2], Xiaodong Li[3], Adri C.T. van Duin[1,2]*

[1] Department of Chemical Engineering, The Pennsylvania State University, University Park, Pennsylvania 16802, United States
[2] Department of Mechanical Engineering, The Pennsylvania State University, University Park, Pennsylvania 16802, United States
[3] Department of Mechanical and Aerospace Engineering, University of Virginia, 122 Engineer's Way, Charlottesville, Virginia 22904, United States
*Corresponding authors. Emails: acv13@psu.edu (Adri C.T. van Duin); xl3p@virginia.edu (Xiaodong Li)

**This PDF file includes:**
**Fig. S1.** An overview of the ReaxFF MD simulations; molecules' geometries and thermal history of simulations
**Fig. S2.** Evolution of N/O-containing species and classification of carbon atom types in graphitic structure
**Fig. S3.** $N_2$ production mechanism by reaction between a PAN chain and an ammonia molecule
**Fig. S4.** $N_2$ production mechanism by reaction between two ammonia molecules

**Fig. S5.** Intramolecular mechanism of $N_2$ production
**Fig. S6.** Intermolecular mechanism of $N_2$ production
**Fig. S7.** RDF for PAN-based CFs and PAN/graphene nanocomposite CFs

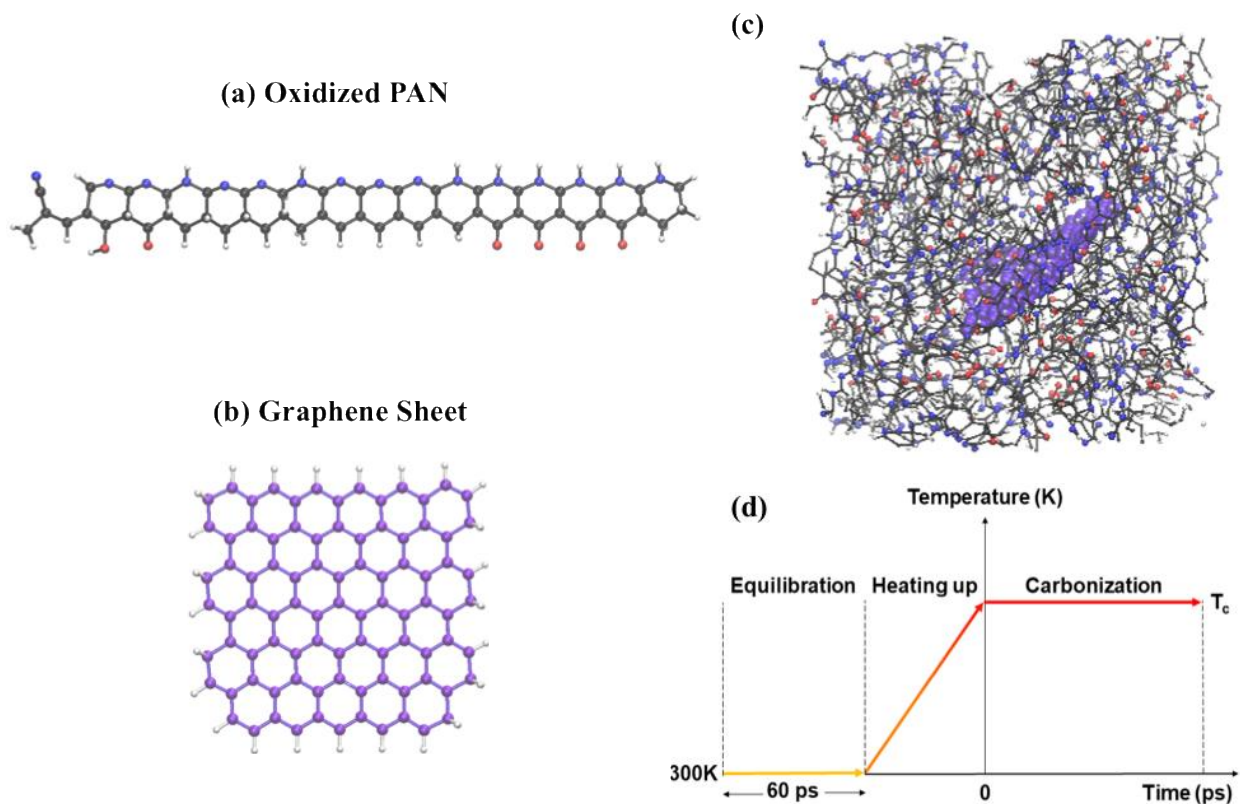

**Fig. S1. An overview of the ReaxFF MD simulations.** (a) oxidized PAN molecular chain, (b) single layer graphene sheet, (c) the simulation box of PAN/graphene nanocomposite including 32 oxidized PAN chains and a single layer graphene sheet after 60ps equilibration at 300K, (d) Simulation profile of temperature versus time. For all analyses we use carbonization part (1 ns) at various carbonization temperatures ($T_c$). Purple spheres represent the initial graphene structure. Carbon, nitrogen, oxygen, and hydrogen atoms are represented in black, blue, red, and white.

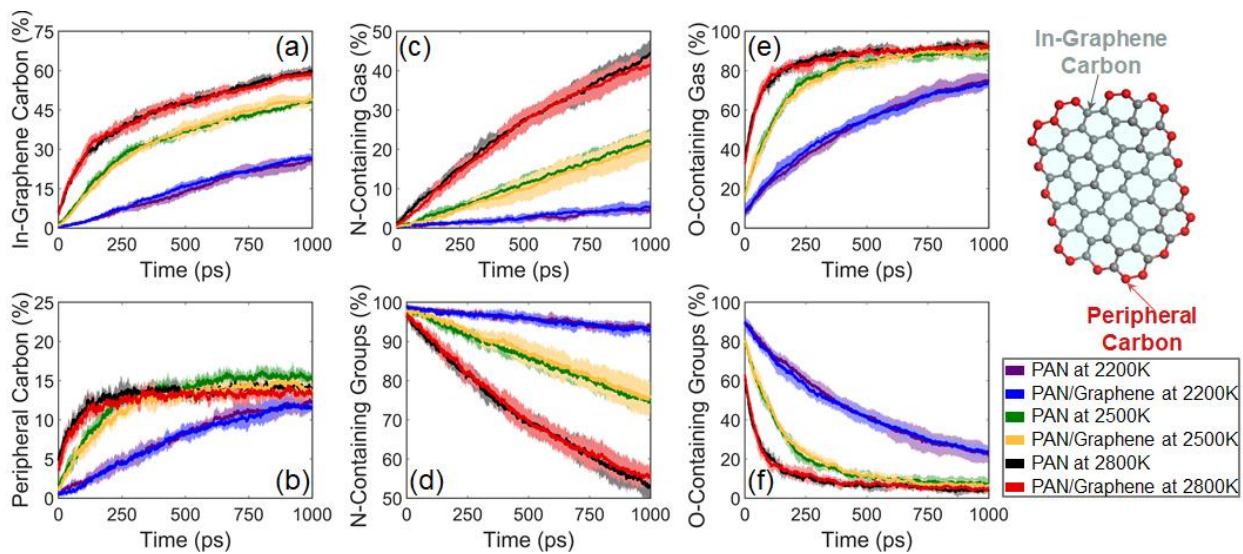

**Fig. S2. Classification of the carbon atoms out of the graphitic network and evolution of N and O-containing species at different carbonization temperatures for PAN-based CFs and PAN/graphene nanocomposite CFs.** (a) In-graphene carbons, (b) Peripheral carbons, (c) N-containing gases, (d) N-containing groups, (e) O-containing gases, (f) O-containing groups, (g) The schematic of in-graphene carbons (gray spheres) and peripheral carbons (red spheres). The N-containing groups consist of the amine groups, imine groups, pyridine-like groups and nitrile groups, and the N-containing gases are mainly $NH_3$ and $N_2$. The O-containing groups include the carbonyl groups, hydroxyl groups and the bridging ether links, and the O-containing gases are mainly $H_2O$, $CO_2$ and CO. nitrogen-containing groups consist of the amine groups, imine groups, pyridine-like groups and nitrile groups, and the nitrogen- containing gases are mainly $NH_3$ and $N_2$. The curves are plotted using the averaged data by 5 samples (dark colors), and the standard deviations are presented by the transparent shadows.

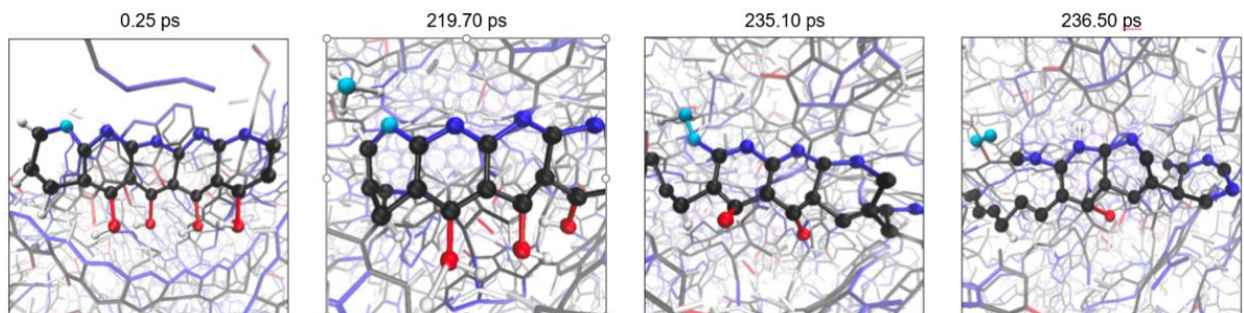

**Fig. S3. N$_2$ production mechanism by reaction between a PAN chain and an ammonia molecule.** Purple spheres represent the initial graphene structure. Carbon, nitrogen, oxygen, and hydrogen atoms are represented in black, blue, red, and white. We color the nitrogen atoms involved in N$_2$ formation as cyan.

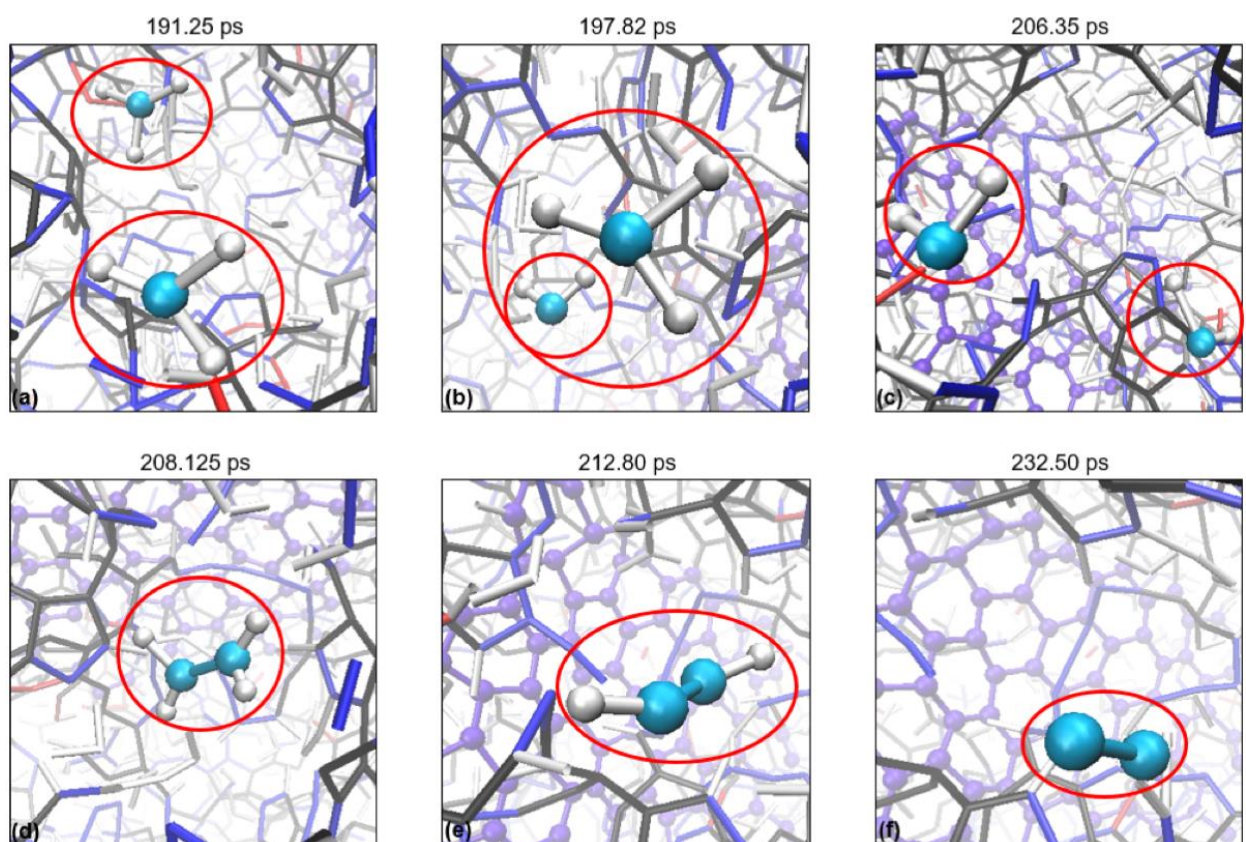

**Fig. S4. N$_2$ production mechanism by reaction between two ammonia molecules.** Purple spheres represent the initial graphene structure. Carbon, nitrogen, oxygen, and hydrogen atoms are represented in black, blue, red, and white. We color the nitrogen atoms involved in N$_2$ formation as cyan.

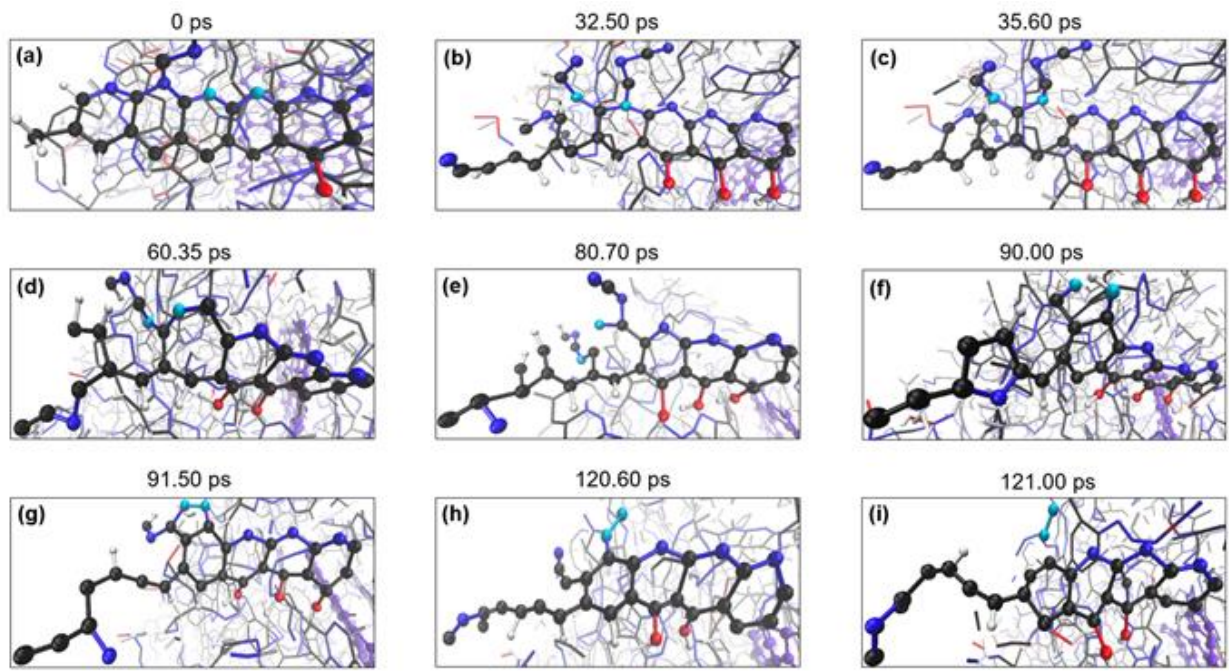

**Fig. S5. Snapshots of $N_2$ production mechanism.** Two nitrogen atoms from one PAN chain form a bond and release an $N_2$ molecule (intramolecular mechanism). Purple spheres represent the initial graphene structure. Carbon, nitrogen, oxygen, and hydrogen atoms are represented in black, blue, red, and white. We color the nitrogen atoms involved in $N_2$ formation as cyan.

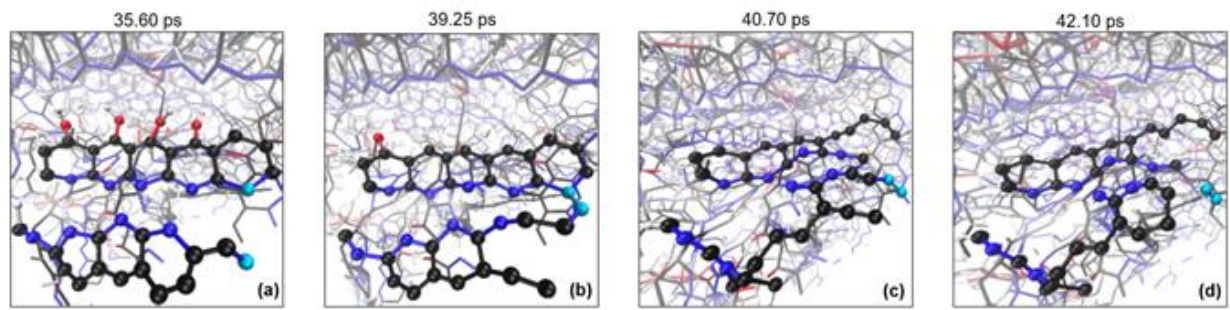

**Fig. S6. Snapshots of $N_2$ production mechanism.** Two nitrogen atoms from two different PAN chains form a bond and release an $N_2$ molecule (intermolecular mechanism) Purple spheres represent the initial graphene structure. Carbon, nitrogen, oxygen, and hydrogen atoms are represented in black, blue, red, and white. We color the nitrogen atoms involved in $N_2$ formation as cyan.

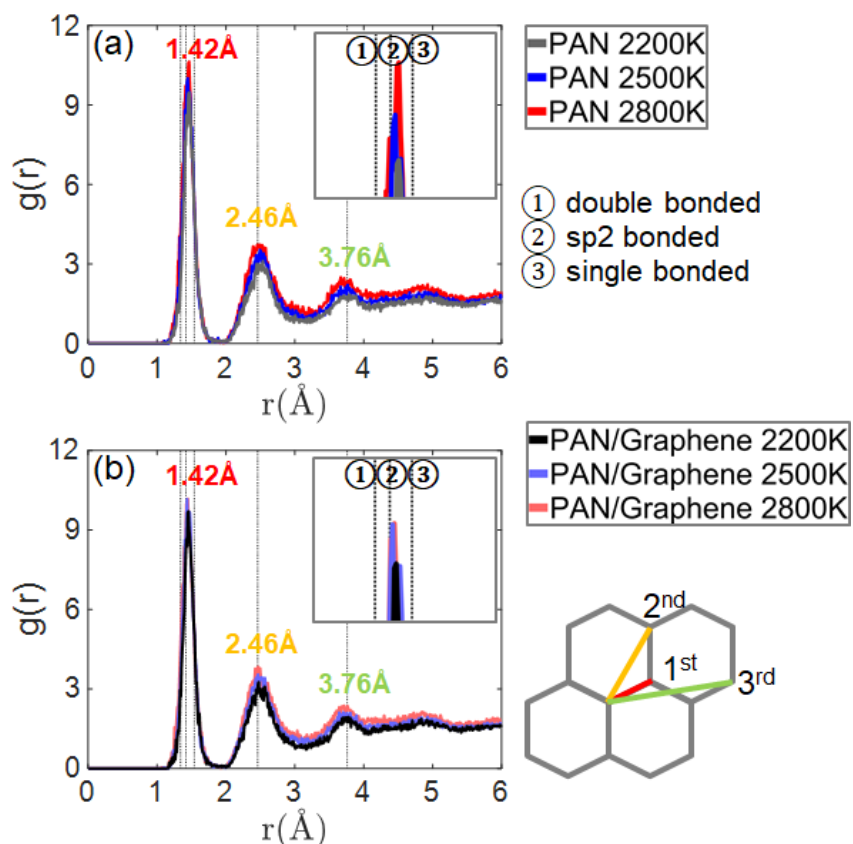

**Fig. S7. Radial distribution function (RDF), denoted as g(r) in arbitrary unit for intensity, at various carbonization temperatures.** (a) C-C pairwise RDF for PAN-based CFs, (b) C-C pairwise RDF for PAN/graphene nanocomposite CFs. The red, yellow, and green values signify the distances r(Å) between the target carbon atom and its 1st, 2nd, and 3rd $sp^2$ bonded neighbor carbon atoms, respectively. The dash lines from left to right on the inset figures correspond to the bond lengths of double bonded, $sp^2$ bonded, and single bonded C-C bonds, which are equal to 1.34 Å, 1.42 Å, and 1.54 Å, respectively. Each RDF curve is plotted using the averaged data by 5 samples.